\documentstyle[aps,preprint,epsf]{revtex}

\begin{document}
\tightenlines
\draft

\title{Vacuum Polarization in QED with World-Line Methods}

\author{W.~ DITTRICH\thanks{
Email:ptidw01@pion06.tphys.physik.uni-tuebingen.de
}
and R.~ SHAISULTANOV\thanks{
Email:shaisul@pion14.tphys.physik.uni-tuebingen.de
}}

\address{Institut f\"{u}r Theoretische Physik, Universit\"{a}t T\"{u}bingen,\\
 Auf der Morgenstelle 14, 72076 T\"{u}bingen, Germany }

\maketitle
\begin{abstract}
Motivated by several recent papers on string-inspired calculations in
QED, we here present our own use of world-line techniques in order to 
calculate the
vacuum polarization and effective action in scalar and spinor QED with
external arbitrary constant electromagnetic field configuration.
\end{abstract}
 \newpage
\narrowtext
\section{Introduction}
String-inspired methods in QFT were initiated by Bern and Kosower
\cite{a}, who applied them to compute one-loop amplitudes in various
field theories. These authors and Strassler \cite{str} then recognized
that some of the well-known vacuum processes in QED and QCD can be
computed rather easily with the aid of one-dimensional path integrals for
relativistic point particles. Similar techniques and results are also
contained in the monograph by Polyakov \cite{pol}. String-inspired
methods, particularly in QED, were then extensively studied in a
series of papers by Schmidt, Schubert and Reuter, cf.,e.g.,\cite{SSR},
where the state of the art is reviewed extensively. 
There are also contributions by McKeon \cite{mk} and various
co-authors who have proved that world-line methods are extremely useful.

In the present article we will try to compactify their work and present
some new representations which will finally permit us to write the
one-loop diagram tied to an
arbitrary number of off-shell photons - together with an applied external
constant electromagnetic field of any configuration - in a so far
unknown manner . This
highly condensed form for the one-loop vacuum process turns out to be
directly applicable to various limiting situations, e.g., turning off
the off-shell photon lines will bring us back to the effective action
of QED which in the low-frequency limit yields the Heisenberg-Euler
Lagrangian. Another process should be photon splitting in a prescribed
constant magnetic field as calculated by Adler and Schubert \cite{as}.

\section{Vacuum Polarization without external Fields - scalar QED.}  

Our starting point is the one loop action, which we write in a path integral
representation,
\begin{equation}
\label{eq:ea1}
\Gamma \left[ A\right]= \displaystyle-i\int ^{\infty }_{0}\frac{dT}{T}{\cal
  N}\int\limits _{x(T)=x(0)}{\cal D}x\: e^{\displaystyle i\int
  ^{T}_{0}d\tau \left[ -\frac{\dot{x}^{2}}{4}-eA\dot{x}-m^{2}\right] },
\end{equation}
where ${\cal N}$ is determined by 
\begin{equation}
\label{eq:ea2}
{\cal N}\int\limits _{x(T)=x(0)}{\cal D}x\: 
e^{\displaystyle-i\int ^{T}_{0}d\tau
  \frac{\dot{x}^{2}}{4}}=\displaystyle-\frac{i}{\left( 4\pi T\right)
  ^{2}}V_{4} .
\end{equation}
In (\ref{eq:ea1}), A denotes a superposition of external plus radiation
field. If we then expand the radiation field in plane waves, $A_{\mu
  }(x)=\sum ^{N}_{i=1}\varepsilon ^{\mu }_{i}e^{ik_{i}x} $, we obtain
\begin{eqnarray}
\Gamma _{N}\left[ k_{1},\varepsilon _{1},\ldots ,k_{N},\varepsilon
 _{N}\right]  & = & -i\left( -ie\right) ^{N}\int ^{\infty
 }_{0}\displaystyle\frac{dT}{T}{\cal N}\int\limits _{x(T)=x(0)}{\cal
 D}x\: e^{\displaystyle i\int
 ^{T}_{0}d\tau \left[ -\frac{\dot{x}^{2}}{4}-eA\dot{x}-m^{2}\right]
 }\cdot \nonumber \\
 &  & \cdot \prod ^{N}_{i=1}\int ^{T}_{0}dt_{i}\varepsilon _{i}\cdot
 \dot{x}(t_{i})e^{\displaystyle ik_{i}\cdot x(t_{i})} \label{eq:ea3}
\end{eqnarray}
Hence, without external field and keeping only terms linear in
$\varepsilon_{i}$,  (\ref{eq:ea3}) becomes
\begin{eqnarray}
\Gamma _{N}=-ie^{N}\int ^{\infty }_{0}
\displaystyle\frac{dT}{T}{\cal N}\int\limits
 _{x(T)=x(0)}{\cal D}x\: e^{\displaystyle i\int ^{T}_{0}d\tau \left[
 -\frac{\dot{x}^{2}}{4}-m^{2}\right] }\left( \prod ^{N}_{i=1}\int
 ^{T}_{0}dt_{i}\right)\cdot &  &  \nonumber \\
\cdot e^{\displaystyle i\sum ^{N}_{j=1}\left( k_{j}\cdot x\left(
 t_{j}\right) -i\varepsilon _{j}\cdot \dot{x}\left( t_{j}\right)
 \right) }\left.\right| _{linear\;in\;each\;\varepsilon_{j} } &  &
 \label{eq:ea4}\;\; .
\end{eqnarray}
Introducing the source for $x^{\mu}$
\begin{equation}
\label{eq:ea5}
J^{\mu }\left( \tau \right) =i\sum ^{N}_{j=1}
\left( k^{\mu }_{j}-\varepsilon ^{\mu }_{j}
\frac{\partial }{\partial t_{j}}\right) \delta \left( \tau -t_{j}\right) ,
\end{equation}
we can rewrite (\ref{eq:ea4}) as
\begin{equation}
\label{eq:ea6}
\Gamma _{N}=-ie^{N}\int ^{\infty }_{0}\frac{dT}{T}e^{-im^{2}T}\left(
  \prod ^{N}_{i=1}\int ^{T}_{0}dt_{i}\right) {\cal N}\int\limits
_{x(T)=x(0)}{\cal D}x\: e^{\displaystyle\frac{i}{4}\int ^{T}_{0}d\tau
  x\frac{d^{2}}{d\tau ^{2}}x}e^{\displaystyle\int ^{T}_{0}J(\tau
  )x(\tau )} .
\end{equation}
The operator $d^2/d\tau^2$, acting on $x(\tau)$ with periodical
boundary condition $x(T)=x(0)$, has zero modes $x_0$. These modes will
be separated from their orthogonal non-zero mode partners by writing
$x(\tau)=x_{0}+\xi(\tau)$ with $\int ^{T }_{0}d\tau
\xi^{\mu}(\tau)=0$, i.e.,$\int ^{T }_{0}d\tau
x^{\mu}(\tau)=x^{\mu}_0$ and $\int {\cal D}x=\int d^{4}x_{0} \int
{\cal D}\xi$.

Since $x(\tau)$ ( and therefore $\xi(\tau)$) is periodic we can write
\begin{equation}
\label{eq:ea7}
x\left( \tau \right) =x_{0}+\sum _{n\neq 0}c_{n}e^{\displaystyle
  \frac{2\pi in}{T}\tau } .
\end{equation}
Now with the use of (\ref{eq:ea2}), equation (\ref{eq:ea6}) 
may be written as
\begin{eqnarray}
\Gamma _{N} & = & -\left( 2\pi \right) ^{4}\delta ^{4}\left( \sum
 ^{N}_{i=1}k_{i}\right) e^{N}\int ^{\infty }_{0}
\displaystyle\frac{dT}{\left( 4\pi
 \right) ^{2}T^{3}}e^{-im^{2}T}\left( \prod ^{N}_{i=1}\int
 ^{T}_{0}dt_{i}\right) \frac{\int {\cal D}\xi e^{\displaystyle\frac{i}{4}\int
 ^{T}_{0}d\tau \xi \frac{d^{2}}{d\tau ^{2}}\xi }
e^{\displaystyle\int ^{T}_{0}J\cdot
 \xi }}{\int {\cal D}\xi e^{\displaystyle\frac{i}{4}\int ^{T}_{0}d\tau \xi
 \frac{d^{2}}{d\tau ^{2}}\xi }} \nonumber \\
 & = & -\left( 2\pi \right) ^{4}\delta ^{4}\left( \sum
 ^{N}_{i=1}k_{i}\right) e^{N}\int ^{\infty }_{0}\frac{dT}{\left( 4\pi
 \right) ^{2}T^{3}}e^{-im^{2}T}\prod ^{N}_{i=1}\int
 ^{T}_{0}dt_{i}\cdot \nonumber \\
&  &\cdot e^{\displaystyle\frac{i}{2}\int ^{T}_{0}d\tau \int ^{T}_{0}d\tau^{
 \prime} J^{\mu }\left( \tau \right) G_{\mu \nu }\left( \tau ,\tau^{
 \prime} \right) J^{\nu }\left( \tau^{\prime} \right) } \label{eq:ea8}\;\;,
\end{eqnarray}
where the Green's function $G_{\mu \nu}$ and its properties are given by
\begin{eqnarray}
G_{\mu \nu }\left( \tau ,\tau^{\prime} \right) =g_{\mu \nu }G\left( \tau
  ,\tau^{\prime} \right) ,\;\;\;\frac{1}{2}\partial ^{2}_{\tau }G\left(
  \tau ,\tau^{\prime} \right) =\delta \left( \tau -\tau^{\prime} \right)
-\frac{1}{T} &  & \nonumber \\
G\left( \tau ,\tau^{\prime} \right) =\left| \tau -\tau^{\prime} \right|
  -\frac{\left( \tau -\tau^{\prime} \right) ^{2}}{T}+const;\;\;G\left(
  \tau ,\tau^{\prime} \right) =G\left( \tau -\tau^{\prime} \right)
  =G\left( \tau^{\prime} ,\tau \right)  &  & \nonumber \\
\partial _{\tau }G\left( \tau ,\tau^{\prime} \right) \equiv
 \dot{G}\left( \tau ,\tau^{\prime} \right) =
sign\left( \tau -\tau^{\prime} \right) -
\frac{2\left( \tau -\tau^{\prime} \right) }{T}
;\;\;\dot{G}\left( \tau ,\tau^{\prime} \right) =
-\dot{G}\left( \tau^{\prime} ,\tau \right).  &  & \label{eq:ea9}
\end{eqnarray}
Note the generic structure expressed in (\ref{eq:ea8}), where particle
and off-shell photons are factorized in such a way that the free
(i.e., sans external field) scalar particle circulating in the loop
becomes multiplied by the exponential term which is solely due to the
photons tied to the loop. With $ J^{\mu }\left( \tau \right)$ given in
(\ref{eq:ea5}) we finally obtain 
\begin{eqnarray}
&&\lefteqn{\Gamma _{N}\left[ k_{1},\varepsilon _{1},\ldots ,k_{N},\varepsilon
  _{N}\right] = -\left( 2\pi \right) ^{4}\delta ^{4}\left( \sum
  ^{N}_{i=1}k_{i}\right) e^{N}\int ^{\infty }_{0}\frac{dT}{\left( 4\pi
  \right) ^{2}T^{3}}e^{-im^{2}T}\prod ^{N}_{i=1}\int ^{T}_{0}dt_{i}} 
\label{eq:ea10} \\
& & \exp {-\frac{i}{2}\sum ^{N}_{i,j=1}\left[ k_{i}\cdot k_{j}G\left(
      t_{i},t_{j}\right) -k_{i}\cdot \varepsilon _{j}\frac{\partial
      }{\partial t_{j}}G\left( t_{i},t_{j}\right) -k_{j}\cdot
    \varepsilon _{i}\frac{\partial }{\partial t_{i}}G\left(
      t_{i},t_{j}\right) +\varepsilon _{i}\cdot \varepsilon
    _{j}\frac{\partial ^{2}}{\partial t_{i}\partial t_{j}}G\left(
      t_{i},t_{j}\right) \right] }\nonumber \; .
\end{eqnarray}
Since $G(\tau,\tau)=0, \dot G(\tau,\tau)$, there are no terms with
$k^{2}_{i}$ and $\varepsilon_{i}\cdot k_i$, i.e., without use of
on-shell conditions.

As an example we just note that for N=2 we obtain, after keeping only
terms linear in $\varepsilon_{i}$,
\begin{eqnarray}
\Gamma _{2}\left[ k_{1},\varepsilon _{1};k_{2},\varepsilon _{2}\right]
=-\left( 2\pi \right) ^{4}\delta ^{4}\left( k_{1}+k_{2}\right)
e^{2}\int ^{\infty }_{0}\frac{dT}{\left( 4\pi T\right)
  ^{2}}e^{-im^{2}T}\int ^{T}_{0}dt_{1}e^{\displaystyle ik^{2}_{1}G\left(
    t_{1}\right) } &  & \nonumber \\
\{\left( k_{1}\cdot \varepsilon _{2}\right) \left( k_{2}\cdot
    \varepsilon _{1}\right) -\left( \varepsilon _{1}\cdot \varepsilon
    _{2}\right) \left( k_{1}\cdot k_{2}\right)\}\dot{G}^{2}\left(
  t_{1}\right) . \label{eq:ea11} &  & 
\end{eqnarray}
This expression is manifestly gauge invariant. Upon using
$G(t_1)=-\frac{\displaystyle t^{2}_{1}}{\displaystyle T}+t_1$,$\dot
G(t_1)=-\frac{\displaystyle 2 t_{1}}{\displaystyle T}+1$ and
    substituting $v=\frac{\displaystyle 2 t_{1}}{\displaystyle T}-1$, we obtain
\begin{eqnarray}
\Gamma _{2}\left[ k_{1},\varepsilon _{1};k_{2},\varepsilon _{2}\right]
=-\left( 2\pi \right) ^{4}\delta ^{4}\left( k_{1}+k_{2}\right)
e^{2}\left[ \left( k_{1}\cdot \varepsilon _{2}\right) \left(
    k_{2}\cdot \varepsilon _{1}\right) -\left( \varepsilon _{1}\cdot
    \varepsilon _{2}\right) \left( k_{1}\cdot k_{2}\right) \right]
\cdot  &  & \nonumber \\
\cdot \int ^{\infty }_{0}\frac{dT}{2\left( 4\pi \right)
  ^{2}T}e^{-im^{2}T}\int ^{1}_{-1}dvv^{2}
e^{\displaystyle ik^{2}_{1}\frac{T}{4}
\left( 1-v^{2}\right) }.\label{eq:ea12} &  & 
\end{eqnarray}
At this point we can make contact with the vacuum polarization digram
in scalar QED:
\begin{equation}
\Gamma _{2}=\left( 2\pi \right) ^{4}\delta ^{4}\left(
  k_{1}+k_{2}\right)\varepsilon_{\mu}\Pi^{\mu \nu}\varepsilon_{\nu},
\end{equation}
where, after renormalization, we obtain for $\Pi_{\mu \nu}=\left(
  g_{\mu \nu }k^{2}-k_{\mu }k_{\nu }\right)\Pi \left( k^{2}\right)$
\begin{equation}
\label{eq:ea13}
\Pi \left( k^{2}\right) =\frac{\alpha }{4\pi }\int ^{1}_{0}dx\left(
  2x-1\right) ^{2}\ln \left[ 1-\frac{k^{2}}{m^{2}}x\left( 1-x\right)
\right] .
\end{equation}
 All these results have been
  reproduced here without the use of operator field theory.

\section{Vacuum Polarization in scalar QED with external Fields.} 

Following the previous discussion on the free-field case we will
briefly describe the effect of an external electromagnetic field on
the single-loop process in scalar QED. Our starting point is a short
review of the results achieved earlier by one of the present authors
\cite{rs}. There it was shown that the action in presence of a
constant electromagnetic background field is given by
\begin{eqnarray}
\Gamma _{N}\left[ k_{1},\varepsilon _{1},\ldots ,k_{N},
\varepsilon _{N}\right] =-\left( 2\pi \right) ^{4}
\delta ^{4}\left( \sum ^{N}_{i=1}k_{i}\right)
 e^{N}\int ^{\infty }_{0}\frac{dT}{\left( 4\pi \right)
 ^{2}T^{3}}e^{-im^{2}T}\cdot  &  & \label{eq:e31}\\
\cdot \prod ^{N}_{i=1}\int ^{T}_{0}dt_{i}\frac{\displaystyle det^{\prime
  1/2}_{P}\left[ \frac{d^{2}}{d\tau ^{2}}\right] }{det^{\prime
  1/2}_{P}\left[ \frac{\displaystyle d^{2}}
{d\tau ^{2}}-2eF\frac{d}{d\tau }\right]
  }e^{\displaystyle \frac{i}{2}\int ^{T}_{0}d\tau \int
  ^{T}_{0}d\tau^{\prime}
 J^{\mu }\left( \tau \right)
 G_{\mu \nu }\left( \tau ,\tau^{\prime} \right)
 J^{\nu }\left( \tau^{\prime} \right) }\nonumber &  & 
\end{eqnarray}
where the Green's function equation is now slightly modified:
\begin{equation}
\label{eq:e32}
\frac{1}{2}\ddot{G}-eF\dot{G}=\delta
 \left( \tau -\tau^{\prime} \right) -\frac{1}{T}.
\end{equation}
$det^{\prime} _{P}$ stands for non-zero modes in the eigenvalue problem
with periodic boundary conditions, just as in the previous
chapter. Also note that our former expression (\ref{eq:ea8}) has
maintained its structure, the only difference being that the former
freely circulating particle is now propagating in the external field
expressed by the field-dependant determinant in (\ref{eq:e31}). As
shown in \cite{SSR} we can write for the ratio of the two determinants
in (\ref{eq:e31})
\begin{equation}
\label{eq:e33}
\frac{det^{\prime 1/2}_{P}\left[ \frac{\displaystyle d^{2}}{d\tau
      ^{2}}\right] }{det^{\prime 1/2}_{P}
\left[ \frac{\displaystyle d^{2}}{d\tau ^{2}}-2eF\frac{d}{d\tau
      }\right] }
=\frac{e^{2}abT^{2}}{\sin \left( ebT\right) \sinh \left( eaT\right) }
\end{equation}
so that the free action of equation (\ref{eq:ea8}) is modified
according to
\begin{eqnarray}
\Gamma _{N}\left[ k_{1},\varepsilon _{1},\ldots ,k_{N},\varepsilon
  _{N}\right] =-\left( 2\pi \right) ^{4}\delta ^{4}\left( \sum
  ^{N}_{i=1}k_{i}\right) e^{N}\int ^{\infty }_{0}\frac{dT}{\left( 4\pi
  \right) ^{2}T}\frac{\displaystyle e^{2}abe^{-im^{2}T}}{\sin \left( ebT\right)
  \sinh \left( eaT\right) }\cdot  &  & \nonumber \\
\cdot \prod ^{N}_{i=1}\int ^{T}_{0}dt_{i}
e^{\displaystyle\frac{i}{2}\int ^{T}_{0}
d\tau \int ^{T}_{0}d\tau^{\prime} J^{\mu }\left( \tau \right)
 G_{\mu \nu }\left( \tau ,\tau^{\prime} \right)
 J^{\nu }\left( \tau^{\prime} \right) } &  & \label{eq:e34}
\end{eqnarray}  
with $G_{\mu \nu }\left( \tau ,\tau^{\prime} \right)$ given as in
\cite{rs}, and a and b are expressed by the invariants
\begin{equation}
a^2=\left( {\cal F}^2 + {\cal J}^2 \right)^{\frac{1}{2}}+{\cal
  F},b^2=\left( {\cal F}^2 + {\cal J}^2 \right)^{\frac{1}{2}}-{\cal F}
;{\cal F}=-\frac{1}{4}F_{\mu \nu }F^{\mu \nu },{\cal
  J}=-\frac{1}{4}F^{*}_{\mu \nu }F^{\mu \nu } \;\; .
\end{equation}
Having recognized similar structures in the free-field case we now
replace (\ref{eq:ea10}) by
\begin{eqnarray}
 &  & \Gamma _{N}\left[ k_{1},\varepsilon _{1},\ldots ,k_{N},\varepsilon
  _{N}\right] =-\left( 2\pi \right) ^{4}\delta ^{4}\left( \sum
  ^{N}_{i=1}k_{i}\right) e^{N}\int ^{\infty }_{0}\frac{dT}{\left( 4\pi
  \right) ^{2}T}\frac{e^{2}abe^{-im^{2}T}}{\sin \left( ebT\right)
  \sinh \left( eaT\right) }\prod ^{N}_{i=1}\int ^{T}_{0}dt_{i}\cdot  
 \nonumber \\
 &  & \lefteqn{\exp {-\frac{i}{2}\sum ^{N}_{i,j=1}\left[ k_{i}G\left(
      t_{i},t_{j}\right) k_{j}-k_{i}\frac{\partial }{\partial
      t_{j}}G\left( t_{i},t_{j}\right) \varepsilon _{j}-\varepsilon
    _{i}\frac{\partial }{\partial t_{i}}G\left( t_{i},t_{j}\right)
    k_{j}+\varepsilon _{i}\frac{\partial ^{2}}{\partial t_{i}\partial
      t_{j}}G\left( t_{i},t_{j}\right) \varepsilon _{j}\right]
      }.}\label{eq:e35} 
\end{eqnarray}
Our next task is to find $\Gamma_2$. However, since the procedure to
arrive at a manifestly gauge invariant expression is straightforward,
we just report our findings:
\begin{eqnarray}
\Gamma _{2}\left[ k_{1},\varepsilon _{1};k_{2},\varepsilon _{2}\right]
=\left( 2\pi \right) ^{4}\delta ^{4}\left( k_{1}+k_{2}\right)
e^{2}\int ^{\infty }_{0}\frac{dTT}{32\pi ^{2}}\int
^{1}_{-1}dv\frac{e^{2}abe^{-im^{2}T}}{\sin \left( ebT\right) \sinh
  \left( eaT\right) }e^{i\Psi }\cdot  &  & \nonumber \\
\cdot \left[ \left( \varepsilon _{1}\rho k\right) 
\left( \varepsilon _{2}\rho k\right) -\left( \varepsilon _{1}\lambda
  k\right)
 \left( \varepsilon _{2}\lambda k\right) -\left( \varepsilon _{1}\rho 
\varepsilon _{2}\right) \left( k\rho k\right) \right]  &  & \label{eq:e36},
\end{eqnarray}
where $k_{1}\equiv k\equiv -k_{2}$ and e.g., $\left( \varepsilon
_{1}\rho k\right) \equiv \varepsilon^{\mu}_{1}\rho_{\mu \nu}
k^{\nu}$, etc. The explicit expressions for $\rho$ and $\lambda$ are
given by
\begin{eqnarray}
\rho =C^{2}\zeta _{3}-B^{2}\zeta _{4},\;  & \lambda =C\zeta
_{1}-B\zeta _{2}, &\label{eq:e37} \\
&& \nonumber\\
{\rm{where}}\; \; \; \; \zeta _{1}=
\frac{\displaystyle\cosh \left( eaT\right) -\cosh
  \left( eaTv\right) }{\sinh \left( eaT\right) }, & \; \; \; \;\zeta
_{2}=\frac{\displaystyle\cos \left( ebT\right) -
\cos \left( ebTv\right) }{\displaystyle \sin
  \left( ebT\right) }, & \nonumber \\
\zeta _{3}=\frac{\displaystyle\sinh \left( eaTv\right)
  }{\displaystyle\sinh \left( eaT\right) }, & \zeta
_{4}=\frac{\displaystyle\sin \left( ebTv\right) }{\displaystyle\sin
  \left( ebT\right) }\;\; . & \nonumber
\end{eqnarray}
The matrices $C_{\mu \nu}$ and $B_{\mu \nu}$ are defined in
\cite{krt}:
\begin{equation}
F_{\mu \nu}=C_{\mu \nu}a+B_{\mu \nu}b , F^{*}_{\mu \nu}=
C_{\mu \nu}b-B_{\mu \nu}a
\end{equation}
and can be expressed in terms of the field strength tensors $F_{\mu
  \nu},F^{*}_{\mu \nu}$ and the invariants a and b. In the special
  situation of parallel electromagnetic fields E,H along the third
  direction, $C_{\mu \nu}$ and $B_{\mu \nu}$ are simply given
  by $C_{\mu \nu}=g^{0}_{\mu} g^{3}_{\nu}-g^{0}_{\nu}g^{3}_{\mu} $,
  $B_{\mu \nu}=g^{2}_{\mu} g^{1}_{\nu}-g^{2}_{\nu}g^{1}_{\mu} $.
However, nowhere in deriving our expression did we employ this special
  field configuration. Therefore, our result is quite general and was
  already considered in \cite{krt}. The same will be true for spinor
  QED, which will be our central result presented in the final
  chapter. Finally we have to define the phase in (\ref{eq:e36}):
\begin{equation}
\Psi =\frac{1}{2}\left[ \frac{\left( kB^{2}k\right) }{eb}\frac{\cos
    \left( ebT\right) -\cos \left( ebTv\right) }{\sin \left(
      ebT\right) }+\frac{\left( kC^{2}k\right) }{ea}\frac{\cosh \left(
      eaT\right) -\cosh \left( eaTv\right) }{\sinh \left( eaT\right)
    }\right]\label{eq:e38} \;\; .
\end{equation}

\section{Vacuum Polarization in spinor QED with external Fields.} 

Here we will start right away with the one-loop action
\begin{equation}
\Gamma \left[ A\right] =\frac{1}{8\pi ^{2}}\int ^{\infty
  }_{0}\frac{dT}{T^{3}}\frac{\int {\cal D}x\: e^{\displaystyle i\int
  ^{T}_{0}d\tau \left[ -\frac{\dot{x}^{2}}{4}-eA\dot{x}-m^{2}\right]
  }}{\int {\cal D}x\: e^{\displaystyle  -i\int ^{T}_{0}d\tau
  \frac{\dot{x}^{2}}{4}}}\frac{\int {\cal D}\psi e^{\displaystyle\int
  ^{T}_{0}d\tau \left[ -\frac{1}{2}\psi _{\mu }\dot{\psi }^{\mu
  }+eF_{\mu \nu }\psi ^{\mu }\psi ^{\nu }\right] }}{\int {\cal D}\psi
  e^{-\displaystyle\frac{1}{2}\int ^{T}_{0}d\tau \psi \dot{\psi
  }}}\label{eq:esp1} \; ,
\end{equation}
where we have introduced the four Grassmann variables
$\psi_{\mu}(\tau)$ which anticommute.

When we use steps similar to those which took us to equation
(\ref{eq:e34}), we arrive again at the background-field one-loop
function multiplied by the off-shell photons tied to the spinor
particle loop:
\begin{eqnarray}
&  &\lefteqn{\Gamma _{N}\left[ k_{1},\varepsilon _{1},\ldots ,k_{N},\varepsilon
  _{N}\right] =\left( 2\pi \right) ^{4}\delta ^{4}\left( \sum
  ^{N}_{i=1}k_{i}\right) e^{N}\int ^{\infty }_{0}\frac{dT}{8\pi
  ^{2}T^{3}}e^{-im^{2}T}\cdot}   \nonumber \\
&  &\cdot \prod ^{N}_{i=1}\int ^{T}_{0}dt_{i}\int d\theta _{i}d\bar{\theta
  }_{i}\frac{det^{\prime 1/2}\left[\displaystyle
 \frac{d^{2}}{d\tau ^{2}}\right]
  }{det^{\prime 1/2}\left[\displaystyle \frac{d^{2}}{d\tau ^{2}}-
2eF\frac{d}{d\tau
      }\right] }\frac{det^{\prime 1/2}\left[ 
\displaystyle\frac{1}{2}\frac{d}{d\tau
      }-eF\right] }{det^{\prime 1/2}\left[
 \displaystyle\frac{1}{2}\frac{d}{d\tau
      }\right] }\cdot   \label{eq:esp2} \\
&  &\cdot \exp {\left\{ \frac{i}{2}\int ^{T}_{0}d\tau \int
  ^{T}_{0}d\tau^{\prime} J^{\mu }\left( \tau \right) G_{\mu \nu
  }\left( \tau ,\tau^{\prime} \right) J^{\nu }\left( \tau^{\prime}
  \right) -\frac{1}{4}\int ^{T}_{0}d\tau \int ^{T}_{0}d\tau^{\prime}
  \eta ^{\mu }\left( \tau \right) \bar{G}_{F\mu \nu }\left( \tau
  ,\tau^{\prime} \right) \eta ^{\nu }\left( \tau^{\prime} \right)
  \right\} }\; ,  \nonumber
\end{eqnarray}
with 
\begin{eqnarray}
\left( \frac{1}{2}\displaystyle\frac{d}{d\tau }-eF\right)
 \bar{G}_{F}\left( \tau ,\tau^{\prime} \right) =
\delta \left( \tau -\tau^{\prime} \right) \nonumber  &  & \\
{\rm{and}}\;\;\;\eta ^{\mu }\left( \tau \right) =\sqrt{2}\sum
^{N}_{j=1}\left( \theta _{j}\varepsilon _{j}^{\mu }-i\bar{\theta
    }_{j}k_{j}^{\mu }\right) 
\delta \left( \tau -t_{j}\right) \; . \label{eq:esp3} &  & 
\end{eqnarray}
As usual, $\theta_{i}$ and $\bar{\theta }_{i}$ denote independent
anticommuting Grassmann variables.

Calculating the determinants yields for the one-loop action with
arbitrary constant electromagnetic field

\begin{eqnarray}
&  &\lefteqn{\Gamma _{N}\left[ k_{1},\varepsilon _{1},\ldots ,k_{N},\varepsilon
  _{N}\right] =\left( 2\pi \right) ^{4}\delta ^{4}\left( \sum
  ^{N}_{i=1}k_{i}\right) e^{N}\int ^{\infty }_{0}\frac{dT}{8\pi
  ^{2}T}e^{-im^{2}T}\cdot}   \nonumber \\
&  &\cdot \prod ^{N}_{i=1}\int ^{T}_{0}dt_{i}\int d\theta _{i}d\bar{\theta
  }_{i}e^{2}ab\frac{\cos \left( ebT\right)
 \cosh \left( eaT\right) }{\sin \left( ebT\right)
  \sinh \left( eaT\right) }\cdot   \label{eq:esp4} \\
&  &\cdot \exp {\left\{ \frac{i}{2}\int ^{T}_{0}d\tau \int
  ^{T}_{0}d\tau^{\prime} J^{\mu }\left( \tau \right) G_{\mu \nu
  }\left( \tau ,\tau^{\prime} \right) J^{\nu }\left( \tau^{\prime}
  \right) -\frac{1}{4}\int ^{T}_{0}d\tau \int ^{T}_{0}d\tau^{\prime}
  \eta ^{\mu }\left( \tau \right) \bar{G}_{F\mu \nu }\left( \tau
  ,\tau^{\prime} \right) \eta ^{\nu }\left( \tau^{\prime} \right)
  \right\} }\; .  \nonumber
\end{eqnarray}
Expressing the sources in the exponential according to equation (\ref{eq:ea5})
and (\ref{eq:esp3}) we obtain
\begin{eqnarray}
\exp \{-\frac{i}{2}\sum ^{N}_{i,j=1}[k_{i}G\left( t_{i},t_{j}\right)
k_{j}-k_{i}\frac{\partial }{\partial t_{j}}G\left( t_{i},t_{j}\right)
\varepsilon _{j}\bar{\theta }_{j}\theta _{j}-\bar{\theta }_{i}\theta
_{i}\varepsilon _{i}\frac{\partial }{\partial t_{i}}G\left(
  t_{i},t_{j}\right) k_{j}+ \nonumber &  & \\
+\bar{\theta }_{i}\theta _{i}\bar{\theta }_{j}
\theta _{j}\varepsilon _{i}\frac{\partial ^{2}}
{\partial t_{i}\partial t_{j}}G\left( t_{i},t_{j}\right)
 \varepsilon _{j}]-\frac{1}{2}\sum ^{N}_{i,j=1}
[\theta _{i}\theta _{j}\varepsilon _{i}\bar{G}_{F}\left(
  t_{i},t_{j}\right) 
\varepsilon _{j}- &  &\label{eq:esp5} \\
-i\theta _{i}\bar{\theta }_{j}\varepsilon _{i}\bar{G}_{F}\left(
  t_{i},t_{j}\right) k_{j}-i\bar{\theta }_{i}\theta
_{j}k_{i}\bar{G}_{F}\left( t_{i},t_{j}\right) \varepsilon
_{j}-\bar{\theta }_{i}\bar{\theta }_{j}k_{i}\bar{G}_{F}\left(
  t_{i},t_{j}\right) k_{j}]\} \; .\nonumber &  & 
\end{eqnarray}
Formula (\ref{eq:esp4}) together with (\ref{eq:esp5}) is our most
general representation for the spin- $\frac{1}{2}$ action with
external fields. Let us now turn to the special case $N=2$. Here we
obtained as a check for the free-field case
\begin{eqnarray}
\Gamma _{2}\left[ k_{1},\varepsilon _{1};k_{2},\varepsilon _{2}\right]
=\left( 2\pi \right) ^{4}\delta ^{4}\left( k_{1}+k_{2}\right)
e^{2}\int ^{\infty }_{0}\frac{dT}{8\pi ^{2}T^{2}}e^{-im^{2}T}\int
^{T}_{0}dt_{1}e^{\displaystyle ik^{2}_{1}G\left( t_{1}\right) }\cdot 
 &  &\nonumber \\
\cdot {\left\{ \left[ \left( k_{1}\cdot \varepsilon _{2}\right) 
\left( k_{2}\cdot \varepsilon _{1}\right) -
\left( \varepsilon _{1}\cdot \varepsilon _{2}\right) 
\left( k_{1}\cdot k_{2}\right) \right] (\dot{G}^{2}-1)\right\} }=
\label{eq:esp6} &  & \\
=-\left( 2\pi \right) ^{4}\delta ^{4}\left( k_{1}+k_{2}\right)
e^{2}\left[ \left( k_{1}\cdot \varepsilon _{2}\right) \left(
    k_{2}\cdot \varepsilon _{1}\right) -\left( \varepsilon _{1}\cdot
    \varepsilon _{2}\right) \left( k_{1}\cdot k_{2}\right) \right]
\cdot  &  & \nonumber \\
\cdot \int ^{\infty }_{0}\frac{dT}{16\pi ^{2}T}e^{-im^{2}T}\int
^{1}_{-1}dv(1-v^{2})e^{\displaystyle ik^{2}_{1}\frac{T}{4}\left(
    1-v^{2}\right) }\label{eq:esp7}\; . &  & 
\end{eqnarray}
Written in the form $\Gamma_{2}=\left( 2\pi \right) ^{4}\delta
^{4}\left( k_{1}+k_{2}\right)\varepsilon _{\mu}\Pi ^{\mu \nu
  }\varepsilon _{\nu}$, this yields the well-known result for the spin-
$\frac{1}{2}$ polarization tensor
\begin{eqnarray}
\Pi ^{\mu \nu }=\left( k^{2}g^{\mu \nu }-k^{\mu }k^{\nu }\right)
 \Pi \left( k^{2}\right), \nonumber &  & \\
{\rm{where}}\;\;\;\; \Pi \left( k^{2}\right) =\frac{2\alpha }{\pi
  }\int ^{1}_{0}dxx\left( 1-x\right) \ln \left[
  1-\frac{k^{2}}{m^{2}}x\left( 1-x\right) \right] \label{eq:esp8}\; . &  & 
\end{eqnarray}

After relatively straightforward calculations which are much easier
and shorter than anything published in the literature we finally end
up with the following form for the spin-
$\frac{1}{2}$ polarization tensor with a prescribed  constant
electromagnetic field of any configuration:
\begin{eqnarray}
\Pi ^{\mu \nu }\left( k\right) =\frac{e^{2}}{16\pi ^{2}}
\int ^{\infty }_{0}dTTe^{-im^{2}T}\frac{e^{2}ab}
{\sin \left( ebT\right) \sinh \left( eaT\right) }
\int ^{+1}_{-1}dve^{i\Psi }\cdot \nonumber &  & \\
\cdot \{{\cal F}_{1}\left( g^{\mu \nu }k^{2}-k^{\mu }
k^{\nu }\right) +{\cal F}_{2}\left[ \left( Ck\right) ^{\mu }
\left( Bk\right) ^{\nu }+\left( Ck\right) ^{\nu }
\left( Bk\right) ^{\mu }\right] +\label{eq:esp9} &  & \\
+{\cal F}_{3}\left[ \left( C^{2}k\right) ^{\mu }
\left( C^{2}k\right) ^{\nu }-\left( C^{2}\right) ^{\mu \nu }\left(
  kC^{2}k\right) \right] 
+\nonumber &  & \\
+{\cal F}_{4}\left[ \left( B^{2}k\right) ^{\mu }
\left( B^{2}k\right) ^{\nu }-\left( B^{2}\right) ^{\mu \nu }\left(
  kB^{2}k\right)
 \right] \}\nonumber &  & 
\end{eqnarray}
Here we introduced
\begin{eqnarray*}
-N_{0}\equiv {\cal F}_{1}=-\cosh \left( eaTv\right) \cos \left(
  ebTv\right)+ \sin \left( ebTv\right) \sinh \left( eaTv\right) \coth 
\left( eaT\right) \cot \left( ebT\right)  &  & \\
N_{3}\equiv {\cal F}_{2}=-\sinh \left( eaTv\right) \sin \left(
  ebTv\right) +
\frac{\left( 1-\cosh \left( eaTv\right) \cosh \left( eaT\right)
  \right) }
{\sinh \left( eaT\right) }\frac{\displaystyle\cos \left( ebTv\right)
  \cos 
\left( ebT\right)-1 }{\sin \left( ebT\right) } &  & \\
N_{1}\equiv {\cal F}_{3}={\cal F}_{1}+2\frac{\cos \left( ebT\right) }
{\sinh ^{2}\left( eaT\right) }\left( \cosh \left( eaT\right) -
\cosh \left( eaTv\right) \right)  &  & \\
N_{2}\equiv {\cal F}_{4}={\cal F}_{1}-2\frac{\cosh \left( eaT\right)
  }{\sin ^{2}\left( ebT\right) }\left( \cos \left( ebT\right) -\cos
  \left( ebTv\right) \right) \; . &  & 
\end{eqnarray*}
Rewriting the tensor structure of the coefficients ${\cal F}_{3}$ and
${\cal F}_{4}$ we obtain
\begin{eqnarray}
\{(\ref{eq:esp9})\}=-N_{0}\left( g^{\mu \nu }k^{2}-k^{\mu }k^{\nu
 }\right)
 +N_{3}\left[ \left( Ck\right) ^{\mu }\left( Bk\right) ^{\nu }+
\left( Ck\right) ^{\nu }\left( Bk\right) ^{\mu }\right] +\nonumber &  & \\
+N_{1}\left( Ck\right) ^{\mu }\left( Ck\right) ^{\nu }-
N_{2}\left( Bk\right) ^{\mu }\left( Bk\right) ^{\nu }\label{eq:esp10}. &  & 
\end{eqnarray}
By introducing the quantities $N_{i}$ we have made contact with the
work of Urrutia \cite{urr}. But while this author treats only parallel
E and H fields we allow for any field direction. Urrutias's result is
reproduced by our formula (\ref{eq:esp9}) together with
(\ref{eq:esp10}) by putting $C_{\mu \nu}=
g^{0}_{\mu} g^{3}_{\nu}-g^{0}_{\nu}g^{3}_{\mu} $,
  $B_{\mu \nu}=g^{2}_{\mu} g^{1}_{\nu}-g^{2}_{\nu}g^{1}_{\mu} $. We
  wanted also mention that Gies \cite{G} has given an alternative but
  much more elaborate, derivation of $\Pi ^{\mu \nu }$. Hence we felt
  that our representation is sufficiently attractive to present it as
  another example that demonstrates how path integral methods together
  with world-line techniques can be put to work.

\section*{Acknowledgments}

The authors profited from discussions with Dr. H. Gies .
This work was supported by Deutsche Forschungsgemeinschaft
under DFG Di 200/5-1.

\end{document}